\documentclass[aps,pre,twocolumn,amsfonts,amsmath,showpacs,a4paper]{revtex4}
\usepackage{color}
\usepackage{epic}
\usepackage{graphicx}
\usepackage{psfrag}
\usepackage{times}

\newcommand{\eg}[1]%                    % exempli gratia
  {{\it e.g.\/}\ifx#1.\else\expandafter#1\fi}
\newcommand{\eq}[1]{(\ref{#1})}         % ref to an equation
\def\eqtwo(#1,#2){(\ref{#1},\ref{#2})}  % ref to two eqns
\newcommand{\etal}[1]%                  % et alia
  {{\it et al.\/}\ifx#1.\else\expandafter#1\fi}
\newcommand{\Fig}[1]{Fig.~\ref{#1}}     % ref to a fig in start of a stce
\newcommand{\Figs}[1]{Figures~\ref{#1}} % ref to figs in a start of a stce
\newcommand{\fig}[1]{fig.~\ref{#1}}     % ref to a fig 
\newcommand{\ie}[1]%                    % id est
  {{\it i.e.\/}\ifx#1.\else\expandafter#1\fi}

\newcommand{\dblfigure}[3]%             % double coulumn figure
  {\begin{figure*}[tbp]#1\caption[]{#2}\label{#3}\end{figure*}}
\newcommand{\sglfigure}[3]%             % single column figure
  {\begin{figure}[tbp]#1\caption[]{#2}\label{#3}\end{figure}}

\renewcommand{\@}{\partial}             % partial differential
  \newcommand{\<}{\langle}              % brac and 
\renewcommand{\>}{\rangle}              %   ket of an inner product
\newcommand{\const}{\mathrm{const}}     % constant
\renewcommand\d{{\mathrm d}}            % ordinary differential
\newcommand{\e}{\mathrm{e}}             % Euler's number
\renewcommand{\Im}[1]{{\rm Im}\left(#1\right)}	% Imaginary part
\renewcommand{\i}{\mathrm{i}}           % imaginary unit
\newcommand{\inner}[2]%                 %  the inner product
  {\left\<#1\, , \,#2\right\>}
\newcommand{\Mx}[1]{%                   % mx and col vec const
\left[\begin{array}{cccccccc}#1\end{array}\right]}
\newcommand{\mx}[1]{\mathbf{#1}}        % mx and col vec vars
\renewcommand{\Re}[1]{{\rm Re}\left(#1\right)}	% Real part
\newcommand{\Real}{\mathbb{R}}          % real set
\newcommand{\T}{^{\mathrm{T}}}          % matrix transposition 

\newcommand{\Ampl}{A}                   % solution amplitude for visualization
\newcommand{\D}{\mx{D}}                 % diffusion matrix
\newcommand{\dfdu}{\@_{\u}\f(\U)}       % kinetics Jacobian
\newcommand\dtime{\dot }                % ordinary time derivative
\newcommand{\F}{F}                      % perturb projections - generic
\newcommand{\f}{\mx{f}}                 % reaction rates
\newcommand{\g}{\mx{g}}                 % spiral perturbation
\newcommand{\h}{\mx{h}}                 % perturbation
\renewcommand{\L}{\mathcal{L}}          % the linearised operator
\newcommand{\Lp}{\L^{+}}                % the adjoint linearised operator
\renewcommand{\O}{\mathcal{O}}          % the approximation oder 
\newcommand{\para}{a}                   % FHN a parameter
\newcommand{\parb}{b}                   % FHN b parameter
\newcommand{\pareps}{\varepsilon}       % FHN epsilon parameter
\newcommand{\R}{{\vec R}}               % R vector (s.w. centre coords) 
\newcommand{\RF}[1]{\mx{W}^{(#1)}}      % left eigenvectors
\renewcommand{\r}{{\vec r}}             % r vector (any pt on the plane)
\newcommand{\Tr}[1]{\mx{V}\ifx#1.\else^{(#1)}\fi} % right eigenvectors
\newcommand{\U}{\mx{U}}                 % unperturbed soln (s.w.)
\renewcommand{\u}{\mx{u}}               % concentration field
\newcommand{\vF}{{\vec{\F}}}            % spatial drift velocity

\newcommand{\Avg}[3]%                   % angular average
  {\langle{#1}\rangle\ifx#2.\else^{(#2)}\fi_{#3}}
\newcommand{\anaTr}[1]%                  % right eigenvectors analytical
  {\breve{\mx{V}}\ifx#1.\else^{(#1)}\fi}
\newcommand{\dr}{\Delta\rho}            % radius step
\newcommand{\dt}{\Delta\theta}          % angle step
\newcommand{\intdist}[1]%               % L2 distance
  {\mathcal{D}_{#1}}
\newcommand{\maxdist}[1]%               % C0 distance
  {\mathcal{D}^{\prime}_{#1}}
\newcommand{\interr}[1]%                % L2 error
  {\mathcal{E}_{#1}}
\newcommand{\maxerr}[1]%                % C0 error
  {\mathcal{E}^{\prime}_{#1}}
\newcommand{\Nr}{N_{\rho}}              % num of steps in radius
\newcommand{\Nt}{N_{\theta}}            % num of steps in angle
\newcommand{\Ntot}{N}                   % tot num of vars of discr pblm
\newcommand{\numlambda}{\hat\lambda}    % num evalues
\newcommand{\nummu}{\hat\mu}            % num evalues
\newcommand{\numomega}{\hat\omega}      % freq numerical
\newcommand{\numRF}[1]%                 % RF numerical
  {\hat{\mx{W}}\ifx#1.\else^{(#1)}\fi}
\newcommand{\numTr}[1]%                 % GM numerical
  {\hat{\mx{V}}\ifx#1.\else^{(#1)}\fi}
\newcommand{\numU}{\hat\U}              % spiral numerical
\newcommand{\Ortana}{O_a}               % orthogonality analytical
\newcommand{\Ortnum}{O_n}               % orthogonality numerical
\newcommand{\rp}{\rho_{\max}}           % disk radius
\newcommand{\B}{{\cal{B}}}            % the whole big disk
\renewcommand{\S}{{\cal{S}}}            % smaller disk

\begin{document}

\title{Computation of the response functions of spiral waves in active media}

\author{I.V.~Biktasheva}
\affiliation{Department of Computer Science, University of Liverpool, 
     Ashton Building, Ashton Street, Liverpool L69 3BX, UK}

\author{D.~Barkley}
\affiliation{Mathematics Institute, University of Warwick, Coventry CV4 7AL, UK}

\author{V. N.~Biktashev}
\affiliation{Department of Mathematical Sciences, University of Liverpool, 
   Mathematics \& Oceanography Building, Peach Street, Liverpool, L69 7ZL, UK}

\author{G.V.~Bordyugov}
\affiliation{former: Department of Computer Science, University of Liverpool, 
     Ashton Building, Ashton Street, Liverpool L69 3BX, UK}

\altaffiliation{Present: The University of Potsdam, Campus Golm,Department of Physics 
and Astronomy (Haus 28), Karl-Liebknecht-Strasse 24/25, 14476 Potsdam, Germany}

\author{A.J.~Foulkes}
\affiliation{Department of Mathematical Sciences, University of Liverpool, 
   Mathematics \& Oceanography Building, Peach Street, Liverpool, L69 7ZL, UK}

\date{\today}

\begin{abstract}
  Rotating spiral waves are a form of self-organization observed in
  spatially extended systems of physical, chemical, and biological
  nature. A small perturbation causes gradual change in spatial
  location of spiral's rotation center and frequency, i.e. drift. The
  response functions (RFs) of a spiral wave are the eigenfunctions of
  the adjoint linearized operator corresponding to the critical
  eigenvalues $\lambda = 0, \pm i\omega$. The RFs describe the spiral's
  sensitivity to small perturbations in the way that a spiral is
  insensitive to small perturbations where its RFs are close to
  zero. The velocity of a spiral's drift is proportional to the 
  convolution of RFs
  with the perturbation. Here we
  develop a regular and generic method of computing the RFs of
  stationary rotating spirals in reaction-diffusion equations. We
  demonstrate the method on the FitzHugh-Nagumo system and 
  also show convergence of the method with
  respect to the computational parameters, i.e.  discretization steps
  and size of the medium.  The obtained RFs are
  localized at the spiral's core.
\end{abstract}

\pacs{%
  02.70.-c, % Computational techniques, simulations
  05.10.-a, % Computational methods in statistical physics and nonlinear dynamics 
  82.40.Bj,% Oscillations, chaos, and bifurcation
  82.40.Ck, % Pattern formation in reactions with diffusion
  87.10.-e % General theory and mathematical aspects 
}

\maketitle

\section{Introduction}

Autowave vortices, or spiral waves in two-dimensions (2D), are types 
of self-organization observed 
in dissipative media of physical \cite{%
  Frisch-etal-1994,%
  Yu-etal-1999,%
  Madore-Freedman-1987,%
  Schulman-Seiden-1986%
}, chemical \cite{%
  Zhabotinsky-Zaikin-1971,%
  Jakubith-etal-1990, %
  Agladze-Steinbock-2000%
}, and biological nature \cite{%
  Allessie-etal-1973,%
  Gorelova-Bures-1983,%
  Alcantara-Monk-1974,%
  Lechleiter-etal-1991,%
  Carey-etal-1978,%
  Murray-etal-1986%
}, where wave propagation is
supported by a source of energy stored in the medium.
The common feature of all these phenomena is that
they can be mathematically described, with various degrees of 
accuracy, by reaction-diffusion partial differential equations,
\begin{equation}
\@_t\u = \f(\u) + \D \nabla^2 \u, \quad 
\u,\f\in\Real^\ell,\; 
\D\in\Real^{\ell\times\ell},\;
\ell\ge2,					\label{RDS}
\end{equation}
where $\u(\r,t)=(u_1,\dots u_{\ell})\T$ 
is a column-vector of the reagent concentrations,
$\f(\u)=(f_1,\dots f_{\ell})\T$ is a column-vector of the reaction rates, 
$\D$ is the matrix of diffusion coefficients, and  
$\r\in\Real^2$ is the vector of coordinates on the plane.

The existence of vortices is not due to singularities in the medium but
is determined only by development from initial conditions.
A rigidly rotating spiral wave solution to the system \eq{RDS} has the form 
\begin{equation}
\tilde \U=\U(\rho (\r-\R),\vartheta (\r-\R) + \omega t - \Phi) ,
                                      \label{SW}
\end{equation}
where 
$\rho(\r-\R),\vartheta(\r-\R)$ are polar coordinates centered at $\R$, 
vector $\R=(X,Y)\T$ 
defines the center of rotation, and $\Phi$ is the initial rotation phase.
For a steady, \ie\ rigidly rotating, 
spiral $\R$ and $\Phi$ are constants. The system of reference 
co-rotating with the spiral's  initial phase and angular velocity $\omega$ around the 
spiral's center of rotation is called the system of reference of the 
spiral. In this system of 
reference, $\R=0$, $ \Phi=0$, and the polar angle is given by $\theta =
\vartheta + \omega t$. In this frame the spiral 
wave solution $\U(\rho,\theta)$ does not depend on time and satisfies the
equation 
\begin{equation}
\f(\U) - \omega \U_\theta + \D \nabla^2 \U = 0 . 	\label{SW-own}
\end{equation}
In this equation, the unknowns are the field $\U(\rho,\theta)$ and the scalar $\omega$.

A slightly perturbed steady spiral wave solution
\begin{equation*}
\tilde \U(\rho,\theta,t)=\U(\rho,\theta)+ \epsilon \g(\rho,\theta,t), \quad 
		\g\in\Real^\ell, \quad 0 < \epsilon \ll 1,
\end{equation*}
substituted in \eq{RDS}, 
at leading order in $\epsilon$,
yields the evolution equation for the perturbation $\g$,
\begin{equation*}
\@_t \g = \dfdu \g 
      - \omega\@_\theta \g + \D\nabla^2 \g.
\end{equation*}

Thus, the linear stability spectrum of a steady spiral  
\begin{equation}
  \L \Tr. = \lambda \Tr.                \label{right-evp}
\end{equation}
 is defined by the linearized operator
\begin{equation}
\L = \D\nabla^2 - \omega\@_\theta + \dfdu.
                                        \label{L}
\end{equation}

The operator $\L$ has critical ($\Re{\lambda}=0$) eigenvalues
\begin{equation}
  \lambda_n=\i n\omega, \quad n=0,\pm1,   \label{lambdas}
\end{equation}
which correspond to eigenfunctions related to equivariance of \eq{RDS}
with respect to translations and rotations, \ie\ ``Goldstone modes'' (GMs)
\cite{%
  Biktashev-1989,%
  Biktashev-1989a,%
  Barkley-1992,%
  Biktashev-Holden-1995%
}
\begin{eqnarray}
\Tr{0} &=&  
  - \@_\theta \U(\rho,\theta), \nonumber\\
\Tr{\pm1} &=& 
% -(\@_x\pm i\@_y)\U(\rho,\theta)\nonumber\\ &=&
  -\frac12 \e^{\mp\i\theta} \left(
    \@_\rho\mp\i\rho^{-1}\@_\theta
  \right) \U(\rho,\theta) .  \label{Goldstone} 
\end{eqnarray}
The stability spectra of steady spiral waves was originally obtained
numerically by Barkley \cite{Barkley-1992}. Subsequently the spectrum was analysed for
infinite and large bounded domains by Sandstede and Scheel~\cite{%
  Sandstede-Scheel-PhysD-2000,%
  Sandstede-Scheel-PhysRevE-2000,%
  Sandstede-Scheel-PhysLett-2001%
} 
with follow-on numerical investigations by Wheeler and Barkley
\cite{Wheeler-Barkley-2006} confirming the large domain behavior
of the stability spectrum.

In a slightly perturbed problem
\begin{equation}
\@_t\u = \f(\u) + \D \nabla^2 \u + \epsilon \h, \quad 
		\h\in\Real^\ell, \quad 0 < \epsilon \ll 1,
                                       \label{RDS_pert}
\end{equation}
where $\epsilon \h(\u,\r,t)$ is some small perturbation, 
spiral waves may drift, \ie change rotational phase and/or center location.
Then, the center of rotation and the initial phase are no longer constants 
but become functions of time, $\R=\R(t)$ and $\Phi=\Phi(t)$.

In linear approximation, assuming that 
\[
  \dtime \R, \; \dtime \Phi = \O (\epsilon),
\]
the drifting spiral wave solution can be represented as 
\begin{equation}
\tilde \U=\U(\rho (\r-\R(t)),\vartheta (\r-\R(t)) + \omega t - \Phi(t)) + \epsilon \g(\r, t),
                                      \label{SW_pert}
\end{equation}
where $\epsilon \g(\r, t) \;$ is a small perturbation of the steady spiral wave solution $\U$.

Then, the solution perturbation $\g$ in the laboratory frame of reference will satisfy the linearized system % \cite{Biktashev-Holden-1995} 

\begin{eqnarray}
(\partial_t &-& \D\nabla^2 - \dfdu) \g
\nonumber\\&&
= \h(\u,\r,t) - \frac{1}{\epsilon}(\dtime \R\cdot\nabla +
\dtime \Phi\,\partial_\theta)\U.
                       \label{g_eqn}             
\end{eqnarray}
The solvalability condition for equation \eq{g_eqn} for $\g \;$, \ie\ Fredholm 
alternative, re-written in the spiral frame
of reference,  requires that the free term must be orthogonal to the kernel of the adjoint operator to $\L$ defined in \eq{L}. 
This leads to the following system of equations for the drift velocities 

\begin{equation}
\dtime\Phi = \epsilon \F_0(\R,t), \quad
\dtime\R=\epsilon\vF_1(\R,t).             \label{ptb}
\end{equation}

Thus, the drift velocities $\dtime \Phi \;$ and $\dtime \R \;$ are determined by the ``forces'' 
$\F_0 \;$ and $\vF_1=\left(\Re{F_1},\Im{F_1}\right)\T$ which, after sliding averaging
(more specifically, central moving average)
over the spiral wave rotation 
period, can be expressed \cite{Biktashev-Holden-1995} as  
\begin{eqnarray}
&&  \F_n(\R,t) =
  \e^{ \i n \Phi }
  \oint\limits_{t-\pi/\omega}^{t+\pi/\omega} 
  \frac{\omega\d\tau}{2\pi}   
  \e^{-\i n\omega\tau} 
\nonumber\\ &&
  \times \inner{
    \RF{n}\left(\rho (\r-\R),\vartheta (\r-\R)+\omega\tau-\Phi\right)
  }{
  \h(\r,\tau)
  } , 
\nonumber\\ &&
  	n=0, \pm1.
  					\label{forces}
\end{eqnarray}
(of course, $F_{-1}=\bar{F_1}$).
Here 
$\inner{\cdot}{\cdot}$ stands for the scalar product in functional space,
\[
  \inner{\mx{w}}{\mx{v}} = \int\limits_{\Real^2}
  \overline{\mx{w}(\r)}\,\T \mx{v}(\r) \,\d^2\r .
\]
The kernels $\RF{n}$ of convolution-type integrals in \eq{forces} 
are the spiral wave's \emph{response functions} (RFs), 
\ie, the critical eigenfunctions 
\begin{equation}
\Lp \RF{n} = \mu_n\RF{n}, 
                      \label{left-evp} 
\end{equation} 
where
\begin{equation}
  \mu_n=-\i\omega n,\quad n=0,\pm1,      \label{mus}
\end{equation}
of the adjoint linearized operator: 
\begin{equation} \Lp = \D\nabla^2 + \omega\@_\theta + \left(\dfdu\right)\T ,  
                               \label{Lp}
\end{equation}
chosen to be biorthogonal 
\begin{equation}
\inner{ \RF{j} }{ \Tr{k} }=\delta_{j,k} ,  		\label{norm}
\end{equation}
to the Goldstone modes \eq{Goldstone}. 
Note that the RFs do not depend on time, \ie\ are functions of the 
coordinates only, in the co-rotating system of reference.

The asymptotic theory just outlined reduces the description of the
smooth dynamics of spiral waves from the system of nonlinear
partial differential equations \eq{RDS} to the system of ordinary
differential equations \eq{ptb}, describing the movement of the
core of the spiral and the shift of its angular velocity.
Several qualitative results in the asymptotic theory of spiral and
scroll dynamics have been obtained without the use of response functions,
\eg.~\cite{%
  Zykov-1987,%
  Davydov-etal-1988,%
  Biktashev-1989a,%
  Keener-Tyson-1990,%
  Keener-Tyson-1991,%
  Davydov-etal-1991,%
  Biktashev-Holden-1994,%
  Biktashev-Holden-1995,%
  Biktashev-1996,%
  Krinsky-etal-1996,%
  Henry-2004%
}.  However, an explicit knowledge of RFs makes possible a
quantitative description, which obviously can be
much more efficient for the understanding and control of spiral
wave dynamics in numerous applications, \eg\ control of re-entry
in the heart.

The asymptotic properties of the RFs at large distances are crucial for convergence of 
the convolution integrals in \eq{forces}.
An early version of the asymptotic theory,
developed by Keener~\cite{Keener-1988} for scroll wave dynamics, 
considered the RFs asymptotically
periodic in the limit $\rho\to\infty$, in much the same way as 
spiral waves are, thus requiring an artifical cut-off procedure to
tackle the divergence of the integrals in \eq{forces} 
following from such an asumption.
  
Based on observations and empirical data of
spiral wave dynamics,
Biktashev~\cite{Biktashev-1989,Biktashev-etal-1994} 
conjectured
that the response functions quickly decay at large $\rho$,
\ie\ are effectively localized.
This conjecture implies that the
integrals in \eq{forces} converge and no cut-off procedure is
required.

To prove existence of the localized responce functions,
Biktasheva \etal\ \cite{Biktasheva-etal-1998} explicitly
computed them in the complex Ginzburg-Landau equation (CGLE) for a
particular set of parameters. Those computations exploited an
additional symmetry present in the CGLE, which permitted the reduction
of the 2D problem to the computation of 1D components. The
computations were verified by numerical convergence of the
method with respect to the space discretisation and the size of
the medium. Following this work, the computed RFs were successfully used for
quantitative prediction of the spiral's resonant drift and drift
due to media
inhomogeneity~\cite{Biktasheva-etal-1999,Biktasheva-2000}. By
explicitly computing the RFs in the CGLE for a broad range of
the model's parameters, Biktasheva and Biktashev~\cite{
  Biktasheva-Biktashev-2001,%
  Biktasheva-Biktashev-2003%
} showed that the RFs are localized for stable spiral wave
solutions and qualitatively change at crossing the
charachteristic lines in the model parameter plane.
 
Recently, there has been a significant theoretical progress in
mathematical treatment of the localization of the response
functions. Sandstede and Scheel~\cite[Corollary
4.6]{Sandstede-Scheel-2004} analytically proved such
localization for one-dimensional wave dislocations, which may be
considered as analogues of a spiral wave in one spatial
dimension. Hopefully this can be extended to two spatial
dimensions, \ie\ to spiral waves.

For cardiac applications, dynamics of
spiral waves in \emph{excitable} media is more important than in 
\emph{oscillatory} media such as the CGLE,
as most cardiac tissues are
excitable. 
These models do not allow 
% meaningful 
reduction to 1D, making quantitatively accurate computation of the response functions more challenging. 
So far, the response functions have been computed 
in the Barkley \cite{Hamm-1997,Henry-Hakim-2002} and FitzHugh-Nagumo
\cite{Biktasheva-etal-2006} models of excitable media.
For the chosen sets of model parameters, the computed RFs appeared
effectively localized in the vicinity of the spiral wave core.  
Hamm~\cite{Hamm-1997} and
Biktasheva~\etal~\cite{Biktasheva-etal-2006} 
calculated RFs on
Cartesian grids, but the accuracy was not sufficient for quantitative
prediction of drift. Hakim and Henry~\cite{Henry-Hakim-2002}
took the advantage of a polar grid and Barkley model 
to compute the spiral wave solution with an
accuracy of
$10^{-8}$ and RFs with accuracy $10^{-6}$
(both in the sense of $l_2$-norm of the residue of the discretized equations) 
leading to
quantitative prediction of drift velocities with about 4\%
accuracy. 
% It is not obvious from the paper \cite{Henry-Hakim-2002}
% how to improve the method further if a higher accuracy is required.

Encouraging as these results are, there is a need for a more 
computationally efficient, accurate and robust
method to compute the response functions of spiral waves in a variety
of excitable media with required accuracy.
The aim of this paper is to present a
method which is superior to previous methods used to 
compute response functions 
and to demonstrate that it works for stationary rotating
spirals in FitzHugh-Nagumo system. We also
demonstrate convergence of the method with respect to the
computational parameters, \ie\ discretization steps and size of the
medium, and show that the method is vastly more
efficient than the
methods used before~\cite{Henry-Hakim-2002,Biktasheva-etal-2006}.
% The obtained response functions are 
% indeed localized at the spiral's core.

\section{Methods}

\subsection{Computations}

To compute the response functions, we use methods similar to those described in
\cite{Barkley-1992,Wheeler-Barkley-2006}. 

The nonlinear problem \eq{SW-own} is considered on a disk $\rho\le\rp$, with
homogeneous Neumann boundary conditions,
$\partial_{\rho}\U(\rp,\theta)=0$.  The fields are discretized on a
regular polar grid $(\rho_j,\theta_k)=(j\dr,k\dt)$ where $0<j\le\Nr$
and $0\le k<\Nt$ plus the center point $\rho=0$. Hence there are
$\Nr\Nt+1$ grid points and correspondingly $\Ntot=\ell(\Nr\Nt+1)$
unknowns and the same number of equations in the discretization of
\eq{SW-own}. For the inner points $j<\Nr$, the $\rho$-derivatives are calculated via second-order
central differences.  The $\theta$-derivatives are calculated using Fornberg's
{\ttfamily weights.f} subroutine \cite{Fornberg-1998} which uses all
$\Nt$ values so, in theory, provides an approximation of
$\theta$-derivatives of the order of $\Nt$. 
The discretization of the
Laplacian at the center point is via the difference between the
average around the innermost circle $\rho=\dr$ and the center point,
and the approximation at $j=\Nr$ takes into account the boundary conditions at
$\rho=\rp$.

The discretized nonlinear steady-state spiral problem \eq{SW-own} is
solved by Newton's method, starting from initial approximations obtained
by interpolation of results of simulations of the time-dependent problem
\eq{RDS} using EZSPIRAL. % The Newton's 
The Newton iterations involve inversion
of the linearized matrix which has a banded structure with the
bandwidth $1+2\ell\Nt$. 
This is achieved by the appropriate ordering of
the unknowns of the discretized problem within the $\Ntot$-dimensional
vector of unknowns, so that the index enumerating components of reagent
vectors from $\Real^{\ell}$ varied fastest, followed by the index
enumerating angular grid points $k\dt$, followed by the index
enumerating the radial grid points $j\dr$. 

The thus posed discretized nonlinear problem inherits the symmetry of
\eq{SW-own} with respect to rotations. To select a unique solution out
of a family of solutions generated by this symmetry, we impose a
``pinning condition'' of the form $U_{\ell_*}(j_*\dr,k_*\dt)=u_*$, where
$\ell_*$, $u_*$ and $j_*$ may be selected arbitrarily and $k_*$
is chosen as the $\theta$-grid point in the $\rho=j_*\dr$ circle that
gives  the $\ell_*$-component value closest to $u_*$ in the initial approximation. Since
$U_{\ell_*}(j_*\dr,k_*\dt)$ is fixed, it is no longer an unknown,
and its place in the $\Real^{\Ntot}$-vector of unknowns is taken by
$\omega$, also to be found from \eq{SW-own}. In this way, the balance of
the unknowns and equations is preserved. As $\omega$ is present in all 
equations, 
the corresponding non-zero column of the linearization matrix 
destroys the bandedness of the matrix. This obstacle is 
overcome by employing the Sherman-Morrison formula
\cite{Numerical-Recipes} to find solutions of the corresponding linear
systems using only banded matrices.
% Newton's 
Newton iterations are performed until the residual in solution of the
discretized version of equation  \eq{SW-own} 
becomes sufficiently small.

The linearized problems \eq{right-evp} and \eq{left-evp} are
considered in the same domain with similar boundary conditions.
The critical eigenvalues and eigenvectors of the discretized operators $\L$ and $\Lp$ are computed
with the help of a complex shift and Cayley transform. 

For a matrix
$\mx{L}$, be it discretization of $\L$ or $\Lp$, the complex
shift is defined as
\[
  \mx{A} = \mx{L} + \i \kappa \mx{I}
\]
and the subsequent Cayley transform as 
\begin{equation}
  \mx{B} = (\xi\mx{I} +  \mx{A})^{-1} (\eta\mx{I} + \mx{A})       \label{mxB}
\end{equation}
where $\kappa$, $\xi$ and $\eta$ are real parameters and $\mx{I}$ is the
identity matrix. If $\lambda$, $\alpha$ and $\beta$ are eigenvalues of
$\mx{L}$, $\mx{A}$
and $\mx{B}$, respectively, this implies 
\[
  \alpha=\lambda+\i\kappa, \qquad \beta=\frac{\eta+\alpha}{\xi+\alpha}.
\]
The selected eigenvalues and eigenvectors of the thus constructed matrices
$\mx{B}$ are then found by the Arnoldi method, 
using ARPACK~\cite{ARPACK}.

 We have used $\xi=0$, $\eta=1$ and
$\kappa=0,\mp \omega$ when seeking, respectively, $\Tr{0,\pm1}$ and
$\RF{0,\mp1}$, where $\omega$ is the solution of the corresponding
nonlinear problem previously obtained. With this choice of $\xi$, $\eta$ and
$\kappa$, the numerical 
eigenvalues $\numlambda$ and $\nummu$ closest to the theoretical critical eigenvalues
\eq{lambdas} and \eq{mus} correspondingly, generate the largest
$|\beta|$. Hence, the Arnoldi method in each case is required to
obtain the eigenvalue with the largest absolute value.

To normalize the eigenvectors, we use the ``analytical'' Goldstone
modes $\anaTr{k}$, obtained by numerical differentiation of the
numerical spiral wave solution $\numU$, namely,
\begin{eqnarray*}
\anaTr{0} &=&  
  - \@_\theta \numU(\rho,\theta), \\
\anaTr{\pm1} &=& 
  -\frac12 \e^{\mp\i\theta} \left(
    \@_\rho\mp\i\rho^{-1}\@_\theta
  \right) \numU(\rho,\theta) ,
\end{eqnarray*}
where differentiation has been implemented using the same discretization 
schemes as used in calculations. 

First, the response functions $\numRF{k}$ computed by ARPACK are
normalized with respect to the ``analytical'' Goldstone
modes $\anaTr{k}$ so that
\[
  \inner{ \numRF{k} }{ \anaTr{k} }=1, \qquad k=0,\pm1,
\]
where numerical integration involved in $\inner{\cdot}{\cdot}$ has been
carried out using the trapezoidal rule. 

Then, the ``numerical'' Goldstone modes $\numTr{k}$ computed by ARPACK 
are normalized with respect to the normalized response functions so that
\[
  \inner{ \numRF{k} }{ \numTr{k} }=1, \qquad k=0,\pm1.
\]

Thus, we finally obtain 
\begin{itemize}
\item a numerical solution for the spiral wave problem \eq{SW-own} together with the angular velocity $\omega$,
\item ``analytical'' Goldstone modes $\anaTr{k}$, 
\item normalized ``numerical'' Goldstone modes $\numTr{k}$, and 
\item normalized response functions $\numRF{k}$.
\end{itemize}

\subsection{Analysis}

\label{sec:anal}

To validate the computed response functions, we have to
demonstrate convergence of the solution with respect to the numerical
approximation parameters such as the size of the medium $\rp$, and
the discretization steps $\dr$ and $\dt$.

First of all, we have to demonstrate convergence  of the computed eigenvalues
of $\numlambda_n$ and $\nummu_n$ to their theoretical values \eq{lambdas}
and \eq{mus}, taking for $\omega$ its numerical approximation $\numomega$ found
by numerical solving the discretized problem \eq{SW-own}. Since the
``theoretical'' value for $\omega$ is not available, we can only check
convergence of $\numomega$ to some limit. 

The accuracy of the ``numerical'' Goldstone modes is quantified by the 
distance between the 
``numerical'' and ``analytical'' Goldstone modes, in $L_2$ norm
\[
  \intdist{j} = \left( \int\limits_{\S}
    \left|\anaTr{j}(\r)-\numTr{j}(\r)\right|^2\,\d^2\r \right)^{1/2}
\]
as well as $C_0$ norm
\[
  \maxdist{j} = \max\limits_{\r\in\S} \left|\anaTr{j}(\r)-\numTr{j}(\r)\right|
\]
over a disk $\S$ of half the radius of the computational domain:
\[
  \S=\{ \r: |\r|\le\rp/2 \} .
\]
The smaller disk is used to exclude the effects of boundary
conditions. The issue is that the exact GM $\anaTr.$ do not satisfy Neumann
boundary conditions whereas $\numTr.$ do, hence there is an inevitable
deviation between them near $\rho=\rp$, which is an artefact of
restricting our problem to a finite domain, and is not indicative of
the accuracy of the computed $\numRF.$, which are expected to be exponentially
small near $\rho=\rp$.

The accuracy of the computed response functions $\numRF.$ could be
tested directly in the same way as the accuracy of the computed
$\numomega$, \ie\ by the numerical convergence to some limit.
This is however, difficult to implement for the numerical
solutions obtained on different grids. Nevertheless, we are able to 
examine the convergence in $\dr$ where coarser grids are subgrids of the
finer grids by restricting the fine-grid solutions to the coarse grid, 
without the need for any interpolation.
Specifically, we calculate 
\[
  \interr{j} = \left( \int\limits_{\B}
    \left|\numRF{j}_{\dr}(\r)-\numRF{j}_{\dr_*}(\r)\right|^2\,\d^2\r \right)^{1/2}
\]
and
\[
  \maxerr{j} = \max\limits_{\r\in\B} \left|\numRF{j}_{\dr}(\r)-\numRF{j}_{\dr_*}(\r)\right|
\]
over the whole computational domain
\[
  \B=\{ \r: |\r|\le\rp \} ,
\]
where $\numRF{j}_{\dr}(\r)$ are the numerical response functions
calculated at the radius step $\dr$ which is an integer multiple of the
minimal radius step $\dr_*$, and the finest numerical response functions
$\numRF{j}_{\dr_*}(\r)$ have been restricted to the coarser grid of
$\numRF{j}_{\dr}(\r)$ of the solution to which they are compared, so the numerical
integration is done over the coarser grid. Note that in the series
with varying $\rp$ and fixed $\dt$ and $\dr$, the coarser grids are
also subgrids of the finer grids, but as the pinning point is
defined via $\rp$, solutions at different $\rp$ are again not
directly comparable to each other so this series is not used 
in this comparison.

We also assess accuracy 
indirectly via the bi-orthogonality between the response functions 
and the Goldstone modes required by \eq{norm}. 
Specifically, we examine
the orthogonality of the RFs 
to the ``analytical'' GMs, 
quantified by
\begin{equation}
  \Ortana=\sum\limits_{j=0,\pm1}\sum\limits_{k=0,\pm1}
  \left|\inner{ \numRF{j} }{ \anaTr{k} } - \delta_{j,k}\right|^2
                                               \label{Ortana}
\end{equation}
and orthogonality of the RFs to the ``numerical'' GMs quantified by
\[
  \Ortnum=\sum\limits_{j=0,\pm1}\sum\limits_{k=0,\pm1}
  \left|\inner{ \numRF{j} }{ \numTr{k} } - \delta_{j,k}\right|^2.
\]
Note, that by construction the diagonal elements of both the ``numerical'' and
``analytical'' bi-orthogonality matrices here are all equal to 1 up to
round-off errors.

The measures $\Ortana$ and $\Ortnum$ require some discussion.
The bi-orthogonality should be exact for exact RFs and GMs.
However, what we calculate are approximations of these functions, subject to
discretization in $\rho$ and $\theta$ and restriction to a finite domain
$\rho\le\rp$. The bi-orthogonality of numerical solutions is therefore not
exact and its deviation from the ideal is an indication of the accuracy of
calculation, and its convergence in $\dr$, $\dt$ and $\rp$ is an indication,
albeit indirect, of the accuracy of the solutions.  

In more detail, if the the matrices representing discretization of
$\L$ and $\Lp$ were transposes of one another, then their
eigenvectors corresponding to different eigenvalues would be exactly
orthogonal in $l_2$, and so a measure of their orthogonality would
not depend on the spatial discretization but only on the accuracy of
the calculation of the eigenvectors by ARPACK. However, $\L$ and
$\Lp$ are conjugate with respect to the scalar product which is
approximated by a discrete inner product with a weight, hence the
matrices of $\L$ and $\Lp$ are not transposed. Moreover, because of
the approximation used for these operators (\eg\ high-order
approximation in $\dt$ vs second-order approximation in $\dr$), the
corresponding matrices are not adjoint of each other with respect to
the weighted $l_2$ either. So, $\Ortnum$ provides a measure of the
consistency of these matrix representations together with the
accuracy with which the eigenvectors are computed with ARPACK.

Moreover, apart from the question of accuracy of finding the eigenvectors
of the discretized operators and accuracy of finding the eigenfunctions 
of the original continuous operators, there remains a question of 
whether the found eigenvectors and eigenfunctions are the ones
that we need, that correspond to $0$ and $\pm i\omega$, 
rather than eigenfunctions corresponding to eigenvalues which happened to be
close to $0$ and $\pm i\omega$~\footnote{
  Close neighbours of the translational eigenmodes are 
  always a possibility in a large enough disk, 
  see \cite{Wheeler-Barkley-2006}.
}. For the GMs, the answer to this question is ensured by
checking the distance $\intdist{j}$; however, this answer is not absolute as the
 comparison
is made only over part of the disk, for reasons discussed above. 
We note, however, that the $\Lp$ eigenfunctions corresponding to the eigenvalues close to but different from $0,\pm i\omega$, are orthogonal to the GMs and for them $\Ortana$ would be not small~\footnote{
  Equation \eq{Ortana} gives $\Ortana=3$ if all nine scalar products vanish;
  however in reality the scalar products of 
  respective GMs and RFs are used for normalization, so in the case
  of wrong RFs, all scalar products would be divided by small
  numbers which may result in rather large values of $\Ortana$. 
}.
Since $\Ortana$ is defined in terms of scalar products with the mode determined
directly from the underlying spiral wave, its smallness
provides the additional assurance that the adjoint eigenfunctions are indeed the RFs
that we are after, not just some adjoint eigenfunctions.

\section{Results}

\subsection{General}

We have tested our method for computing the response functions in the case of
the FitzHugh-Nagumo model, $\ell=2$,
\begin{eqnarray*}
f_1 & = & \pareps^{-1}(u_1-u_1^3/3-u_2), \nonumber \\
f_2 & = & \pareps(u_1-\para u_2 + \parb),
\end{eqnarray*}
$\D=\Mx{1&0\\0&0}$, with parameters $\para=0.5$,
$\parb=0.68$, $\pareps=0.3$. 
For pinning, we have used $\ell_*=2$, $u_*=0.1$ and $j_*=\Nr/2$. 
% Newton's
Newton
iterations have been performed until the Euclidean ($l_2$) norm of the
residual in the discretized nonlinear equation falls below
$10^{-8}$. For comparison, we have also run cases, discussed later in 
\fig{comp}, in which iterations continue until the
norm of the residual no longer decreases (typically such norms were
below $10^{-9}$ down to $10^{-13}$). 
The tolerance in ARPACK's
routines {\ttfamily znaupd} and {\ttfamily zneupd} has been set to the
default ``machine epsilon''. For the Krylov subspace dimensionality we
have tried 3 and 10, with no perceptible difference in either the
numerical results.

%%%%%%%%%%%%%%%%%%%%%%%%%%%%%%%%%%%%%%%%%%%%%
\dblfigure{
  \includegraphics{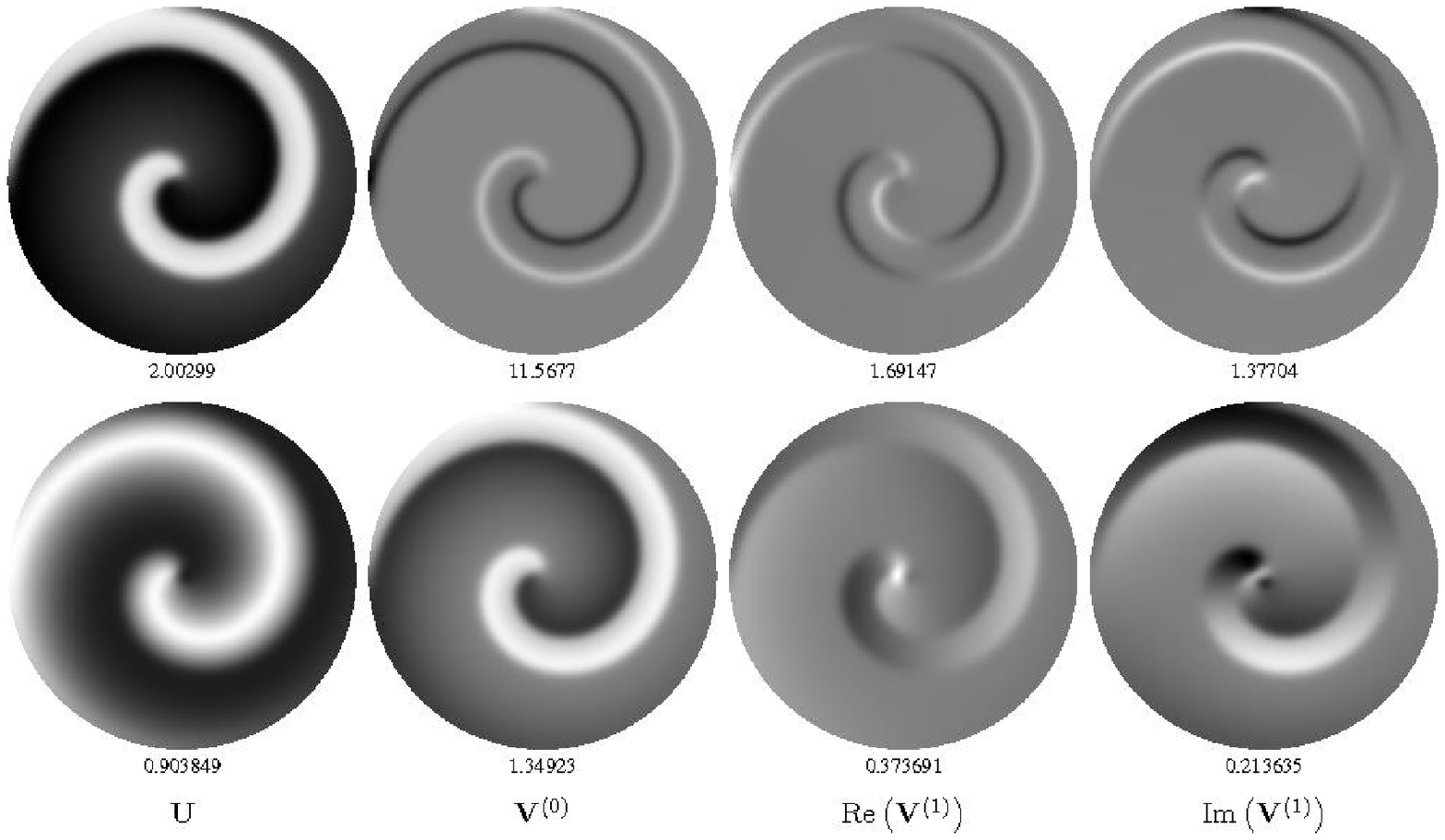}
}{
  Solutions of the nonlinear problem \eq{SW-own} and the linearized problem
  \eqtwo(right-evp,L), \ie\ the Goldstone modes, at the ``best'' parameters,
  $\rp=25$, $\Nr=1280$, $\Nt=64$, 
  as density plots. 
  Numbers under the density plots are their amplitudes $\Ampl$: 
  white of the plot corresponds to the value $\Ampl$ and black corresponds
  to the value $-\Ampl$ of the designated field.
  Upper row: 1st components, lower row: 2d components.
}{gmpics}
%%%%%%%%%%%%%%%%%%%%%%%%%%%%%%%%%%%%%%%%%%%%%

%%%%%%%%%%%%%%%%%%%%%%%%%%%%%%%%%%%%%%%%%%%%%
\dblfigure{
  \includegraphics{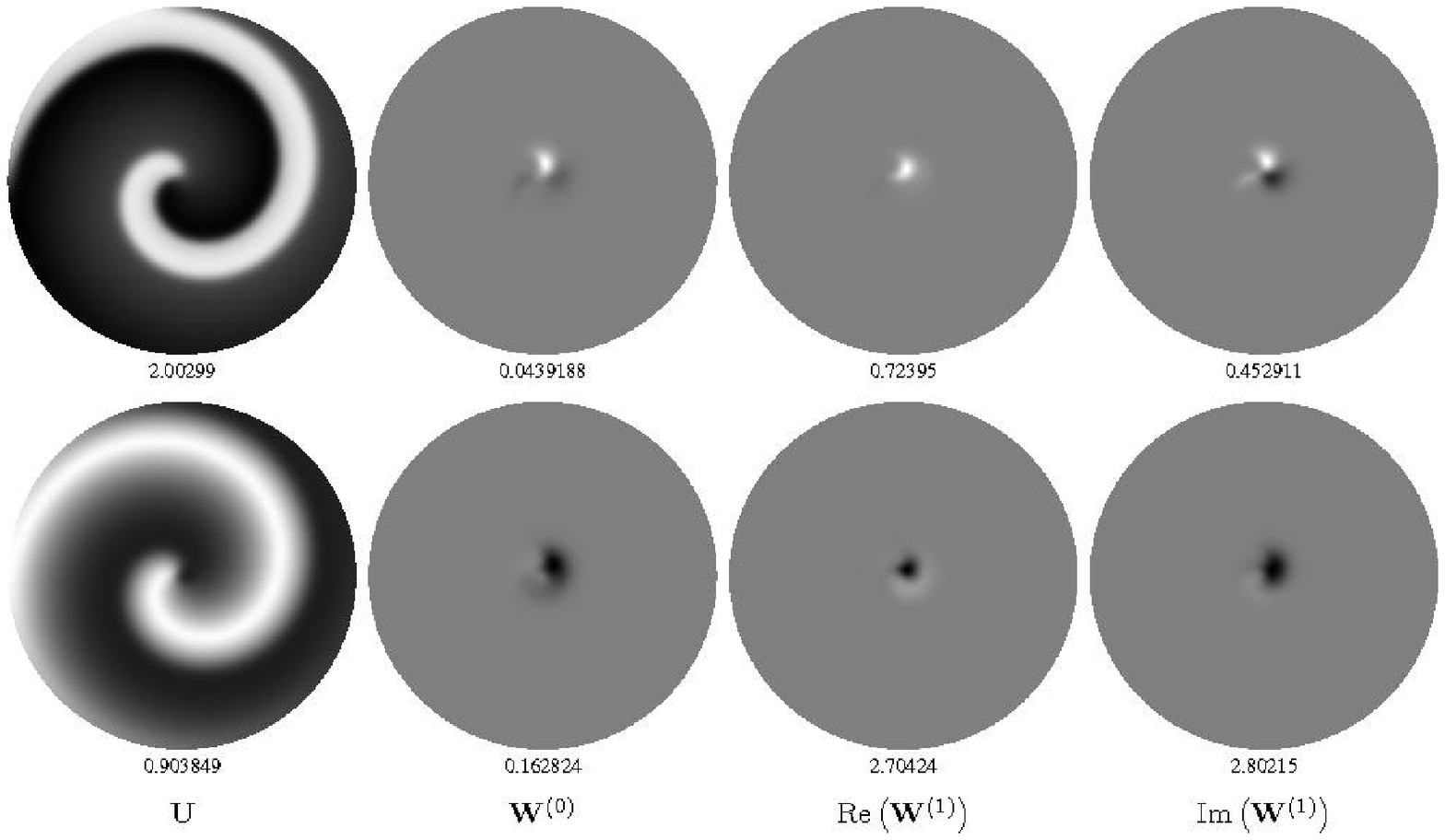}
}{
  Same visualization as in \fig{gmpics}, for the adjoint linearized problem
  \eqtwo(left-evp,Lp), \ie\ the response functions. 
}{rfpics}
%%%%%%%%%%%%%%%%%%%%%%%%%%%%%%%%%%%%%%%%%%%%%

%%%%%%%%%%%%%%%%%%%%%%%%%%%%%%%%%%%%%%%%%%%%%
\dblfigure{
  \includegraphics{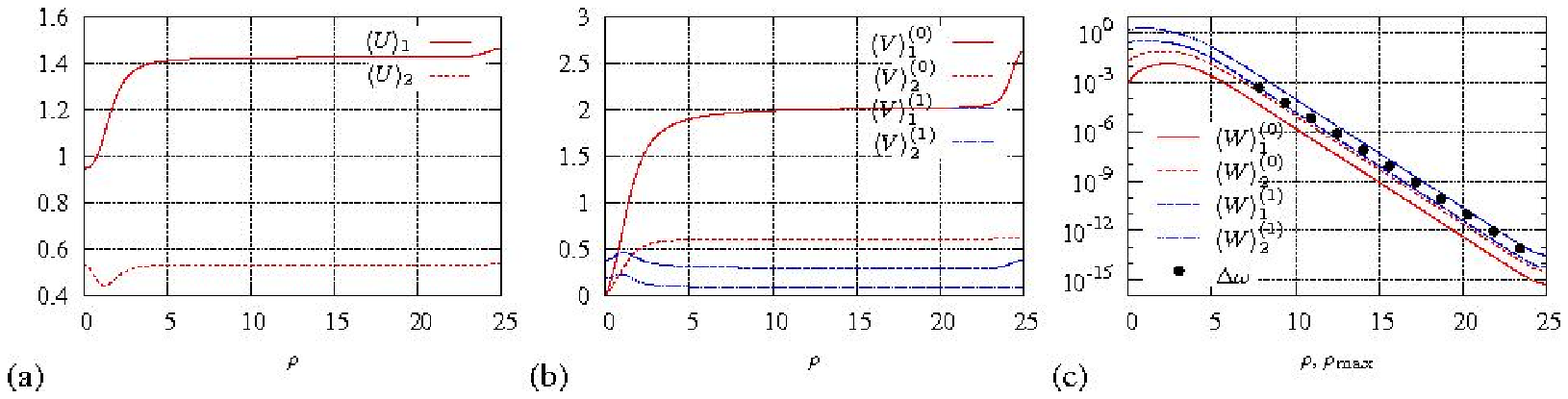}
}{
  Radial dependence of the angle-averaged solutions for the spiral
wave (a), Goldstone modes (b) and response functions (c).  In (c), the
dependence of $\Delta\omega(\rp)=\numomega(\rp)-\numomega(25)$ 
is shown for comparison, where
$\numomega(\rp)$ is the numerically
found spiral angular velocity in the disk of given radius $\rp$. 
}{decay}
%%%%%%%%%%%%%%%%%%%%%%%%%%%%%%%%%%%%%%%%%%%%%

Before discussing the performance of our numerical techniques, we briefly 
present typical solutions.
\Figs{gmpics} and \ref{rfpics} 
illustrate the spiral wave solution and the GMs and
RFs for $\rp=25$, $\Nr=1280$ and $\Nt=64$. This solution is taken as 
the best achievable given memory restrictions (4Gb
of real memory).
The angular velocity for it was found to be 
$\numomega\approx0.5819341748776017$.
For the GMs and RFs, we show the $n=0$ and $n=1$ modes only, since the 
calculated $n=-1$ modes are almost exactly the 
complex conjugates of the $n=1$ modes, which of course they should be.

One can see that the GMs $\numTr.$ are indeed proportional to
corresponding derivatives of the spiral wave solution $\numU$, and
that the RFs $\numRF.$ are localized in a small region of the spiral
tip and are indistinguishable from zero outside that region.

The character of the RFs' decay with distance is illustrated in more
detail in \fig{decay}. We plot the
angle-averaged values of the solutions, defined as
\[
  \Avg{X}{n}{i}(\rho) = \frac{1}{2\pi} \oint \hat{X}_i^{(n)}(\rho,\theta)
  \,\d\theta,
\]
for $X=U,V$ and $W$. 
Note the difference in the behavior of $\Avg{U}{n}{i}$ and 
$\Avg{V}{n}{i}$ on one hand and $\Avg{W}{n}{i}$ on the other hand.
In the
semilogarithmic (linear for horizontal axis, logarithmic
for vertical axis)
coordinates of \fig{decay}(c) the graphs of
$\Avg{W}{n}{i}(\rho)$ are straight for a large range of $\rho$,
not too close to 0 or $\rp=25$, and for several decades of magnitude of
$\Avg{W}{n}{i}$. 
% This is an evidence of the expected exponentialcharacter of decay.
This shows clearly the expected exponential localization of the RFs.
For comparison, we also show
convergence of $\numomega=\numomega(\rp)$ in a disk as a function of the
disk radius $\rp$.  Theory \cite{Hagan-1982, Biktashev-1989-inNW, Biktasheva-Biktashev-2001, Sandstede-private} predicts that the
$\Avg{W}{n}{i}(\rho)$ and $\Delta\omega(\rp)=\numomega(\rp)-\numomega(\infty)$
dependencies should both be decaying 
exponentials with the same characteristic exponent; this
agrees well with the numerical results shown in \fig{decay}(c).

Sandstede and Scheel~\cite{%
  Sandstede-Scheel-PhysRevE-2000,%
  Sandstede-Scheel-PhysLett-2001%
} have computed exponential decay/increase rates of eigenfunctions of
periodic wavetrains in one spatial dimension. A similar technique
should, in principle, also work for the adjoint eigenfunctions. Knowning the
asymptotic wavelength of the spiral wave, this can be used
to predict the exponential decay rates of the RFs of spiral
waves. As can be seen from the results of Wheeler and Barkley~\cite{Wheeler-Barkley-2006},
although such correspondence between 1D and 2D calculations 
can be established, the accuracy of decay rate estimates for two-dimensional
eigenfunctions achieved in this way is insufficient for a meaningful estimate of the
accuracy of those eigenfunctions.

%%%%%%%%%%%%%%%%%%%%%%%%%%%%%%%%%%%%%%%%%%%%%
\dblfigure{
  \includegraphics{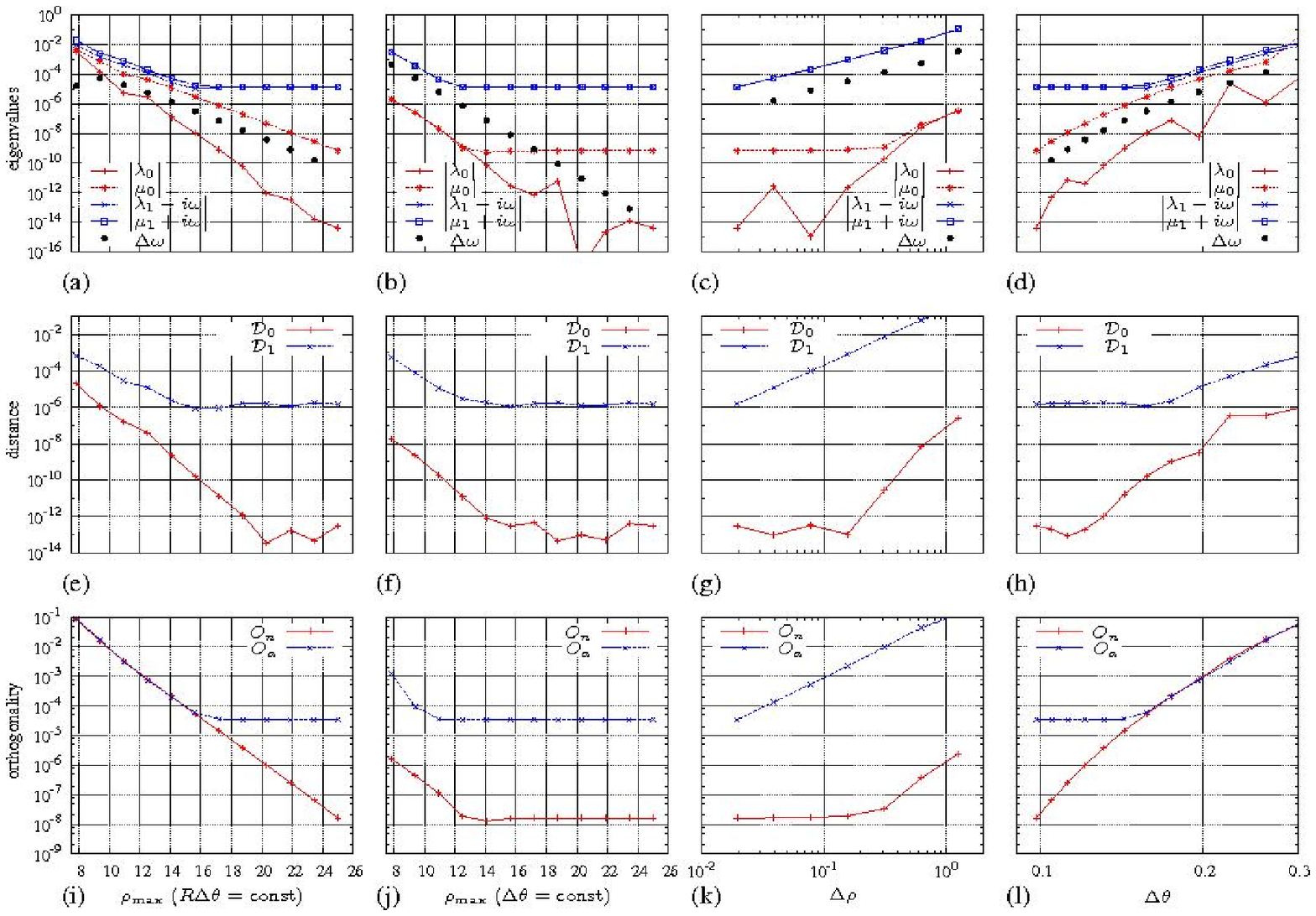}
}{
  Convergence in numerical parameters: of deviation of the numerical
  eigenvalues from theoretical (upper row), of $L_2$ distance between numerical
  and theoretical eigenfunctions (second row) and of orthogonality,
  \ie\ Frobenius norm of the difference of the matrix
  of scalar products of eigenfunctions and adjoint eigenfunctions from
  the unity matrix (third row), all in logarithmic scales,
  as dependencies of disk radius (first
  and second columns, linear scale), radius
  discretization step (third column, logarithmic scale) and
  polar angle discretization step
  (fourth column, logarithmic scale).
  In the first column, $\rp$ is changed while
  the values of $\dr$ and
  $R\dt$ are kept constant.  
  In the second column, $\rp$ is changed while $\dr$
  and $\dt$ are kept constant. 
}{convergence}
%%%%%%%%%%%%%%%%%%%%%%%%%%%%%%%%%%%%%%%%%%%%%

\subsection{Convergence}

We now turn to the main results of our study. 
Convergence of the method has been tested by changing one of the three
numerical approximation
parameters $\rp$, $\Nr$ and $\Nt$ while keeping the other two at the fixed
values set by the ``best example''. More specifically, while changing
$\rp$, we consider two variants: one with fixed $\Nt$, and one with changing
$\Nt$ so that the combination $\rp\dt$, which is the size of the
outermost computational cells in the angular direction, remains
constant. 

\Fig{convergence} illustrates the results of the study, where the four
columns correspond to different series of calculations, and the three
rows correspond to the three different methods of assessing the
accuracy: closeness of the eigenvalues to the theoretical values,
distance between ``numerical'' and ``analytical'' GMs and
orthogonality between non-dual RFs and GMs. The scales of $\Delta \rho$, $\dt$
and the error estimates are logarithmic, and the scales of $\rp$ are
linear. Here shown is the distance between the ``numerical'' and ``analytical'' 
Goldstone modes in $L_2$ norm, the distance in $C_0$ norm looks similar.

A typical feature on many of the curves is a ``knee''-shape, when the
measure of the error decreases as $\rp$ grows or $\dt$ or $\Delta \rho$
decrease, but only until a certain point, beyond which it 
reaches a plateau. This behavior is expected and explicable. 
The calculation error is affected by many factors, and if the factor
varied in a particular series becomes negligible, then the error
remains at a constant level determined by fixed values of other factors.

The position of the ``knees'' on the curves indicates that the
accuracy of the rotational ($n=0$) modes would be improved if $\dt$ were
further decreased (there are no knees on the curves corresponding to the rotational modes, red online, in the fourth,
\ie\ rightmost column), whereas the limiting parameter for the
translational ($n=1$) modes is $\dr$ (there are no knees on the curves corresponding to the tanslational modes, blue online, in
the third column). The analysis of the first two columns is more
complicated. The errors estimates at the maximal $\rp$ are similar in
both columns as they correspond to the same ``best'' spiral. These
limit values are achieved, \ie\ plateaux are observed, at much smaller
$\rp$ values if $\dt=\const$, than if $\rp\dt=\const$.  This is
because reduction of $\rp$ at fixed $\dt$ produces an additional
improvement of approximation due to the angular discretization.  When
$\rp\dt$ is kept fixed, as in the first column, the dependence of the
solution on the disk radius is without this extra benefit.

The rates of convergence with respect to parameters can be assessed by the
slopes of the curves above the knees before they plateau. In some cases the
data is somewhat irregular, primarily at parameters corresponding to lower
values of error estimates. This is not unexpected and we attribute it to
incomplete convergence of the iterative procedures (see below).  On the whole,
the slopes can be determined clearly from these plots.

The constant slope in the first (leftmost) and the second columns
corresponds to the exponential convergence with $\rp$. The constant
slope in the third column corresponds to power-law convergence, and
the typical slope is 2. This is well seen on the curves for
translational modes, blue online, and not well on the curves for rotational
modes, red online, which are very small anyway. 
Slope 2 in the third column is to be
expected as our discretization is second-order in $\dr$ in all
cases. The curves in the fourth (rightmost) column are convex, which
is consistent with the fact that the order of approximation is $\Nt$,
which varies along the curve as $\dt$ varies, since $\Nt=2\pi/\dt$, so
the slope is bigger for smaller $\dt$. 
In other words, the high order of the Fornberg approximation of the
$\theta$ derivatives implies the convergence in $\dt$ is faster than
any fixed power.

%%%%%%%%%%%%%%%%%%%%%%%%%%%%%%%%%%%%%%%%%%%%%
\dblfigure{
  \includegraphics{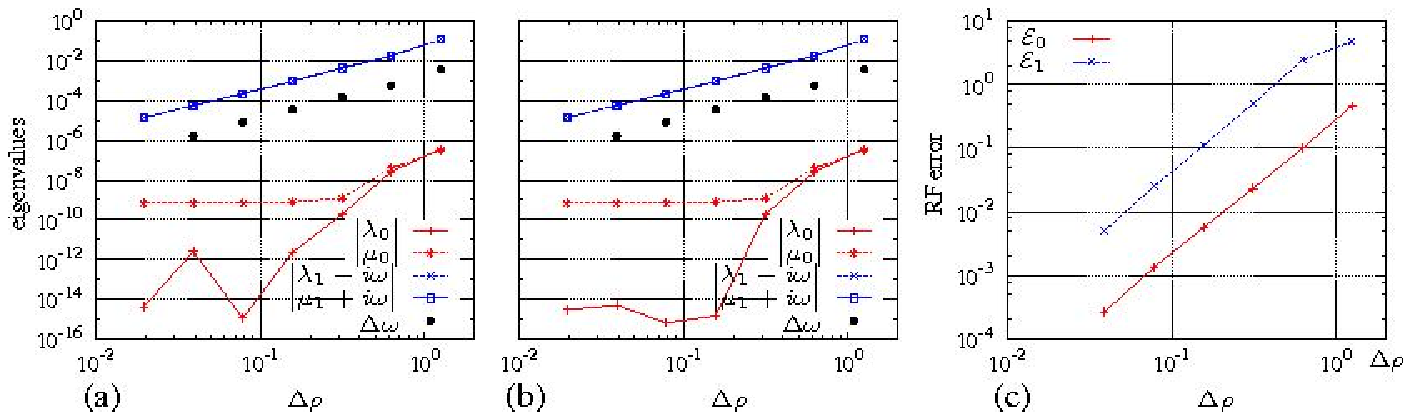}
}{
  (a,b)
  Effect of the accuracy of the unperturbed spiral wave solution on the convergence:
  (a) Newton-iteration tolerance $10^{-8}$.
  (b) Newton iterations until the norm of the residual
  stopped decreasing. 
  (c) Convergence of the response functions in $\dr$. 
}{comp}
%%%%%%%%%%%%%%%%%%%%%%%%%%%%%%%%%%%%%%%%%%%%%

The irregular shape of some of the curves in
\fig{convergence} at very low values of the error estimates is related to
the accuracy of finding the spiral solution and is ulitmately affected by the
precision of floating point computations. 
Note that all calculations in \fig{convergence}
have been performed with %Newton's 
a tolerance of $10^{-8}$ for Newton iterations of the spiral wave and some of
the curves fall as low as $10^{-15}$ \ie\ close to machine
epsilon. A change in the % Newton's iteration 
tolerance of the Newton iteration reduces irregularities in the curves
at low values, as shown in \fig{comp}(a,b).

Finally, \fig{comp}(c) illustrates convergence of
numerical RFs $\numRF{0,1}$ as $\dr\to0$, calculated as the
$L_2$-distance $\interr{0,1}$ between the solutions at a given
resultion $\dr$ and the ``best'' solution calculated at the smallest
$\dr_*=25/1280$.  As explained in the Sec.~\ref{sec:anal}, this
comparsion has been restricted to the series of calculations with varying
$\dr$, where grids at lower resolutions were subgrids of those with
higher resolutions. The graphs of $C_0$ distances $\maxerr{0,1}$
looked similar and are not shown here.

\section{Discussion}

The main result of this paper is
a general, robust method for obtaining response functions
for rigidly rotating spiral waves in excitable media
with required accuracy.

We have tested the method 
on the FitzHugh-Nagumo model
and we have studied the convergence of 
spiral wave solutions and eigenfunctions, 
both the Goldstone modes and the
response functions,
with respect to the numerical approximation parameters $\rp$, $\Nr$
and $\Nt$. The rates of convergence are found to agree with the order of
approximation and indicate the accuracy with which solutions can be found for
particular numerical parameters.

The slowest (second-order) convergence is, as expected, in the
parameter $\Nr$. Thus in a typical situation, an improvement of
accuracy requires, other things being equal, an increase of $\Nr$,
with associated increase in memory and time demands. Thus, the
most promising avenue of further development of the method is via increase of
the approximation order of the radial derivatives.
This is, of course, subject to usual caveat that the degree of approximation
should be consistent with the actual smoothness of the solutions.

The method used here to solve the eigenvalue problems for operators $\mx{L}$
relies on successive application of 
% linear 
transformations of $\mx{L}$ applied to 
a sequence of vectors, alternating with Gram-Schmidt
orthogonalization. These are typical ideas, also used
in \cite{Henry-Hakim-2002,Biktasheva-etal-2006}. The difference is
that in \cite{Henry-Hakim-2002,Biktasheva-etal-2006}, the 
linear transformations were polynomial functions of 
$\mx{L}$ whereas we use rational functions of
$\mx{L}$.  
The polynomial iterations used in
\cite{Henry-Hakim-2002,Biktasheva-etal-2006} were in fact equivalent to
solving a Cauchy problem for equation
$\d{\u}/\d{t}=\mx{L}\u$ by the explicit Euler method.  Therefore, those
methods require a large number of iterations, and
convergence speed of the iterations depends on the smallness of the
absolute difference of the real parts of the eigenvalues of interest
compared to those of other eigenvalues. 
One requires at least $\O(10^5)$ and typically 
$\O(10^6)$ sparse matrix-vector
multiplications to achieve the desired solutions to the eigenvalue problem using
such an approach. 

In contrast, with the
complex shift and inversion of $\mx{L}$ used in this paper, the
convergence speed of the iterations depends on the smallness of the
distance of the eigenvalues from their theoretical values used in
the complex shift, compared to the distance to other eigenvalues. Hence
the number of iterations required is very small, typically $\O(10)$.
More specifically, with Krylov subspace dimensionality 3, the number of matrix
multiplications with matrix $\mx{B}$ of \eq{mxB} 
did not exceed 7 per one eigenpair; with Krylov subspace
dimensionality 10, this number rose to 10.  
The price to pay for this acceleration is the necessity to solve 
large systems of linear equations.
However, the key observation is that since the linear system is fixed, 
it needs to be factorized only once, for a given complex shift, and used for 
all iterations. Multiplication by matrix $\mx{B}$ is achieved with only
inexpensive back/forward solves. 
Moreover, due to the way we ordered the unknowns in the discretized problem,
the sparcity of matrix $\mx{B}$ does not depend on the order of
approximation of $\theta$-derivatives. Hence, we are able to employ high-order
approximations requiring far fewer points in the $\theta$ direction
for the same accuracy as the second-order finite difference
discretization used in \cite{Henry-Hakim-2002}, thereby further improving
the efficiency of our method. 

Discounting the factorization step, each
iteration, which involves multiplication by $\mx{B}$, 
is comparable to multiplications by $\mx{L}$. 
In practice we find that the factorization itself does not require
more than the equivalent of four to six actions of $\mx{B}$.
On a MacPro with 3\,GHz Intel processor, 
the factorization step
takes \eg\ about 7.5\,sec for the grid $\Nr=1280$, $\Nt=64$, and 
0.67\,sec for the grid $\Nr=640$, $\Nt=32$; the computation times 
per $\mx{B}$-multiplication were 1.23~sec and 0.17\,sec
respectively.  

The comparison of our present method
with \cite{Biktasheva-etal-2006} is unequivocal: matrix inverses
were not used there, and 
it was admitted already in \cite{Biktasheva-etal-2006}
that the resulting accuracy of solutions was severely limited.
While direct accuracy and timing comparisons with \cite{Henry-Hakim-2002}
would be most convincing, that code is not publicly available. However, for
reasons already noted, on any given polar grid, the method we report is more
accurate due to the angular discretization and considerably faster in
floating-point operations.

The computed response functions are localized in the vicinity of the
spiral wave tip and exponentially decay with distance from it. This
localization ensures convergence of the convolution integral in
\eq{forces} in an unbounded domain.

The eigenvectors of the linearized operator, \ie\ Goldstone modes and
of its adjoint, \ie\ the response functions have been computed
using the same technique, so the qualitatively different behavior of
these solutions at large $\rho$ is not a numerical artefact, as it was
not in any way assumed in the numerical method.

Although the method has been used here to compute the response functions in
the FitzHugh-Nagumo model, none of the details of the method depends
on any specifics of the particular
reaction kinetics and should be widely applicable to the computation of
response functions of rigidly rotating waves in any other model of excitable
tissue, as long as its right-hand sides are continuously differentiable so the linarized theory is applicable.
Moreover, the method can also be extended in a straightforward way to include
additional effects, such as the effect of uniform twist along scroll waves
with linear filaments in three
dimensions~\cite{%
  Biktashev-1989a,%
  Margerit-Barkley-2001,%
  Henry-Hakim-2002%
}.

\section*{Acknowledgement}
This study has been supported in part by EPSRC grants EP/D074789/1 and EP/D074746/1.

%\bibliography{rf}

\end{document}